# Huge *ac* magnetoresistance in $La_{0.7}Sr_{0.3}MnO_3$ in sub-kilo gauss magnetic fields

A. Rebello, V. B. Naik and R. Mahendiran[1]

Department of Physics and NUS Nanoscience & Nanotechnology Initiative

(NUSNNI), Faculty of Science, National University of Singapore,

2 Science Drive 3, Singapore -117542, Singapore

**Abstract**

We report dynamical magnetotransport in a ferromagnetic metallic oxide, $La_{0.7}Sr_{0.3}MnO_3$ using the *ac* impedance technique. The temperature dependence of the *ac* resistance($R$) and the inductive reactance ($X$) of the complex impedance ($Z = R+jX$) under different *dc* bias magnetic fields ($\mu_0 H_{dc}$ = 0-1 kOe) were studied for different frequencies $f$ = 0.1 to 5 MHz of alternating current. The zero field $R$, which decreases smoothly around the Curie temperature $T_C$ for $f$ = 100 kHz, transforms into a peak for $f$ = 0.5-5 MHz. The peak decreases in amplitude, broadens and shifts downward in temperature as the bias field increases. A huge *ac* magnetoresistance ($\Delta R/R$ = 45 % at $f$ = 2 MHz) in a field of $\mu_0 H_{dc}$ = 1 kOe is found and we attribute it to the magnetic field- induced enhancement in the skin depth and concomitant suppression of magnetic fluctuations near $T_C$. Our study suggests that radio frequency magnetotransport provides an alternative strategy to enhance the magnetoresistance and probe the spin-charge coupling in manganites.

---

[1] Corresponding author – phyrm@nus.edu.sg

Electrical response of conduction electrons to microwave radiation in normal metals has been a topic of considerable interest in solid state physics.[1] Experiments such as cyclotron resonance and anomalous skin effect, which study the dynamics of electrons in the GHz electromagnetic fields and static magnetic fields, have been exploited to map Fermi surfaces in clean metals since 1930's.[2] Applications of microwave absorption to study surface impedance and dynamics of flux lattices in superconductors,[3] high frequency excitation in charge density wave systems,[4] ferromagnetic and antiferromagnetic resonance in magnetic materials are well known.[5] Recently, there has been a renaissance of interest in radio frequency (*rf*) electrical transport in conducting ferromagnets following the discovery of giant magnetoimpedance (GMI) effect in some amorphous ferromagnetic alloys in the frequency range *f* = 0.1-10 MHz and due to their potential applications in novel magnetic sensors and devices.[6,7] The GMI effect refers to a large change in the *ac* impedance $Z(\mu_0 H_{dc}, f) = R(\mu_0 H_{dc}, f) + jX(\mu_0 H_{dc}, f)$ of a ferromagnetic conductor carrying a radio frequency (*f*) current in response to an external *dc* magnetic field ($\mu_0 H_{dc}$) applied parallel to the direction of the current. The low-field magnetoimpedance ($-\Delta Z/Z$ = 40-100 % in a field of $\mu_0 H_{dc}$ = 10-100 Oe) is much larger than the *dc* magnetoresistance which is typically 10 % in a high field of $\mu_0 H_{dc}$ = 20-50 kOe. The low-field sensitivity of magnetoimpedance is claimed to be comparable to superior to flux gate and giant magnetoresistance sensors.[7] The GMI effect is generally attributed to changes in the magnetic penetration depth ($\delta$) upon the application of a *dc* magnetic field. The flow of an alternating current (*ac*) in a ferromagnetic wire creates an *ac* magnetic field ($\mu_0 H_{ac}$) in the transverse direction to the current, which induces a transverse *ac* magnetization ($M_{ac}$). Application of a *dc* bias field decreases the transverse permeability $\mu_t = M_{ac}/H_{ac}$, which leads to an increase in the skin depth according to the relation $\delta = (2\rho/\omega\mu_0\mu_t)^{1/2}$ and causes decrease in the *ac* impedance, since $Z = (1+j)\rho/\delta$ where $\rho$ is the *dc* resistivity. However, not all ferromagnetic materials, but only those with low resistivity and transverse magnetic domain structures with high transverse permeability can show the GMI effect.

In contrast to the amorphous ferromagnets, crystalline ferromagnetic manganites such as $La_{0.67}Sr_{0.33}MnO_3$ show nearly 100 % magnetoresistance ("colossal magnetoresistance") in a magnetic field of $\mu_0H_{dc}$ = 50-70 kOe.[8] The domain structures of manganites are hardly known and so is the influence of *ac* current on the resistance. The resistivity of manganites is a factor of 10 to 100 times higher than amorphous metallic alloys. Moreover, origins of electrical conduction and ferromagnetism in manganites are very different from the amorphous ferromagnets. It is natural to ask whether manganites can show the giant magnetoimpedance effect at low magnetic fields. While the *ac* impedance in the absence of an external magnetic field was studied in the past to investigate dynamics of charge localization or inter or intra-grain electrical conduction in insulating or semiconducting manganites,[9] only a very few studies on radio frequency (*rf*) magnetotransport in the metallic ferromagnetic manganites are currently available. A considerable magnetoimpedance effect ($\Delta Z/Z$ = 5-15 % under $\mu_0H_{dc}$ = 5-8 kOe) in $La_{1-x}A_xMnO_3$ (A = Sr, Ba, x ≈ 0.25-0.3) in the frequency range *f* = 1-12 MHz were reported while passing *rf* current directly through the sample[10] or even higher value while measuring magnetoimpedance of an induction coil enclosing the sample.[11] However, these studies were mostly confined to room temperature or temperature dependence of the impedance at a fixed magnetic field.[12] A systematic study of the *ac* impedance at close intervals of magnetic fields will be helpful to understand the origin of magnetoimpedance in these manganites, which underlines the motivation behind the present work.

In this communication, we report our work on the *rf* magnetoimpedance ($f$ = 100 kHz to 5 MHz) under low magnetic fields ($\mu_0 H_{dc}$ < 1 kOe) in a simple ferromagnet $La_{0.7}Sr_{0.3}MnO_3$ which is considered to be a canonical example of a double exchange ferromagnet with a Curie temperature of $T_C$ = 365±2 K.[8] This particular composition does not suffer from the mesoscopic or microscopic phase segregation which is often encountered in narrow band width manganites such as $Pr_{0.7}Ca_{0.3}MnO_3$.[13] The static Jahn-Teller effect is rather weak in this metallic composition (x = 0.3)[14] and small angle neutron scattering experiments are not in strong supportive of the existence of magnetic polarons unlike in $La_{0.7}Ca_{0.3}MnO_3$.[15] Hence, $La_{0.7}Sr_{0.3}MnO_3$ is an ideal compound to investigate the *rf* magnetotransport.

Polycrystalline $La_{0.7}Sr_{0.3}MnO_3$ sample was prepared by the standard solid state route and characterized by X-ray diffraction, *dc* magnetization and *ac* susceptibility measurements. Four probe resistive ($R$) and inductive ($X = \omega L$) components of the complex *ac* impedance, $Z = R+jX$, of a rectangular sample of dimension 10.1 mm x 3.6 mm x 2.2 mm with 5.3 mm distance between the voltage probes were measured using an Agilent 4285A impedance analyzer with an *ac* excitation of voltage amplitude, $V_{rms}$ = 20 mV in the frequency range $f$ = 100 kHz to 5 MHz. The sample was mounted in a specially designed multifunctional probe wired with coaxial cables and open/short and load corrections were done at room temperature to reduce errors in the measurements due to cable capacitance and inductance. The temperature dependence of $Z$ under different $dc$ bias magnetic fields applied perpendicular to the length of the sample was measured using a commercial superconducting cryostat (PPMS, Quantum Design Inc, USA).

Figure 1 shows the temperature dependence of the *ac* resistance ($R$) (on the left scale) and inductive reactance ($X$) (on the right scale) of $La_{0.7}Sr_{0.3}MnO_3$ in zero magnetic field, $\mu_0 H_{dc}$ = 0 Oe and for $f$ = 100 kHz. The temperature dependence of the $dc$ resistivity is identical to the $R$(0 Oe, 100 kHz) data. The resistivity is 9.04 mΩ cm at 400 K and 5.34 mΩ cm at 300 K for $f$ = 100 kHz. The *ac* resistance shows a metallic behavior ($dR/dT > 0$) from 400 K down to 300 K, but a rapid decrease in $R$ occurs between 370 K and 350 K. The downward pointed arrow indicates the Curie temperature $T_C$ obtained from the *ac* susceptibility measurement (not shown here). The rapid decrease in $R$ is accompanied by a steep increase in the zero-field inductive reactance ($X = \omega L$) around $T = T_C$ = 363 K. Since $L \propto \mu$, where $\mu$ is the low-frequency initial permeability, the steep increase of $X$ demarcates the paramagnetic region above $T_C$ from the ferromagnetic region below $T_C$. Thus the electrical impedance measurement, compared to $dc$ electrical transport, has an added advantage of probing charge transport and magnetism simultaneously.

The temperature dependence of the *ac* resistance ($R$) on the left scale and the reactance ($X$) on the right scale for $f$ = 100 kHz under different $dc$ bias magnetic fields are shown in Figs. 2(a) and 2(b). While the value of $R$ at $f = 100$ kHz is hardly affected by the external bias magnetic fields, the amplitude of $X$ decreases with increasing strength of the magnetic field below the $T_C$. Figures 2(c) and 2(d) show the $R$ and $X$ for $f = 500$ kHz. In contrast to the $R$(0 Oe, 100 kHz) data, the $R$(0 Oe, 500 kHz) shows a steep increase around $T_C$ followed by a maximum just below $T_C$ and then decreases with lowering temperature. The applied *dc* magnetic field has no impact on the resistance above $T_C$. However, the maximum in $R$ decreases, broadens and shifts towards lower temperatures with increasing strength of $H_{dc}$. The maximum in $R$(1 kOe, 500 kHz) is completely suppressed under $\mu_0 H_{dc} = 1$ kOe and the behavior is restored to the temperature dependence of $R$ at $f = 100$ kHz or *dc* resistivity in the zero field. The extraordinary sensitivity of the resistance maximum to sub kilo gauss magnetic fields shown is very interesting for practical applications. The reactance at $f = 500$ kHz is also very sensitive to the *dc* bias field as can be seen in fig. 2(d).

Figures 3(a) and 3(b) show the behavior of $R$ and $X$ respectively at $f = 1$ MHz. The $R$(0 Oe, 1 MHz) shows a steep increase around $T_C$ and a maximum close to it. While the amplitude of the maximum decreases with $\mu_0 H_{dc}$, it is not completely eliminated under $\mu_0 H_{dc} = 1$ kOe unlike for $f = 500$ kHz. The behavior of $X$ under different *dc* bias fields is similar to the low frequency data. The temperature dependence of $R$ and $X$ for $f = 2$ MHz (Figs. 3(c) and 3(d)) under different *dc* magnetic fields is identical to the $R$ and $X$ for $f = 1$ MHz, but with a difference in the magnitudes. It can be seen that as the frequency of the alternating current increases, magnitudes of $R$ and $X$ in zero field increase in the entire temperature range investigated due to skin effect.

Next we show the temperature dependence of the *ac* magnetoresistance, $\Delta R/R$ = [$R$(0 Oe) - $R(\mu_0 H_{dc})$]/$R$(0 Oe)×100 at $f$ = 2 MHz in Fig. 4(a). As $T$ decreases from 400 K, the $\Delta R/R$ at $\mu_0 H_{dc}$ = 300 Oe is negligible in the paramagnetic state, but shows an abrupt increase at $T_C$ followed by a sharp peak close to $T_C$. The magnitude of the peak increases from 20 % at $\mu_0 H_{dc}$ = 300 Oe to a remarkable value of 45 % at $\mu_0 H_{dc}$ = 1 kOe. As the strength of the *dc* bias field increases, the peak gets broadened due to the enhancement of $\Delta R/R$ for $T \ll T_C$. The observed values of the *ac* magnetoresistance are indeed remarkable since a magnetic field of at least $\mu_0 H_{dc}$ = 60-70 kOe is necessary to induce a *dc* magnetoresistance of the same value. The inset of Fig. 4(a) shows the peak value of $\Delta R/R$ at $\mu_0 H_{dc}$ = 1 kOe as a function of frequency. The magnitude of the peak *ac* magnetoresistance increases with frequency, goes through a maximum of 45 % at 2 MHz before it decreases to 40 % at $f$ = 5 MHz. Figure 4(b) shows the magnetoreactance, $\Delta X/X$ = [$X$(0 Oe) - $X(\mu_0 H_{dc})$]/$X$(0 Oe)×100 at $f$ = 2 MHz, which exhibits a peak just below $T_C$, but the amplitude of the peak ($\Delta X/X$ = 14 % at $\mu_0 H_{dc}$ = 1 kOe) is smaller than $\Delta R/R$. However, $\Delta X/X$ has the maximum value at the lowest frequency (= 42 % at $f$ = 0.1 MHz) and it decreases monotonically with increasing frequency to about 7 % at $f$ = 5 MHz as can be seen in the inset of Fig. 4(b). The magnetoimpedance, $\Delta Z/Z$ = [$Z$(0 Oe) - $Z(\mu_0 H_{dc})$]/$Z$(0 Oe)×100 at $f$ = 2 MHz reaches a maximum value of 15 % at $\mu_0 H_{dc}$ = 1 kOe and $T$ = 355 K. The $\Delta Z/Z$ increases from 3 % at $f$ = 0.1 MHz to a maximum of 24 % at 500 kHz and then decreases with further increase in frequency as shown in the inset Fig. 4(c). It has to be mentioned that the magnitude of the observed radio frequency magnetoimpedance is smaller than ~70 % microwave magnetoimpedance at $\mu_0 H_{dc} = 600$ Oe reported in single crystals of $La_{0.7}Sr_{0.3}MnO_3$.[16] However, mechanisms of magnetization dynamics at microwave and radio frequencies are different. While magnetization primarily proceeds by spin rotation in microwave range, domain wall movement is important at lower frequencies.

The observed peak in the *ac* magnetoresistance close to $T_C$ is opposite to the behavior of the grain boundary related magnetoresistance in polycrystalline manganites. In polycrystalline ferromagnetic manganites, the low field magnetoresistance arises from the enhanced spin polarized tunneling of $e_g$ holes between misaligned ferromagnetic grains through spin disordered grain boundaries.[17] This grain boundary magnetoresistance is largest at 4.2 K and decreases to zero at $T_C$. Intrinsic magnetoresistance of intra-grain origin is dominant close to the magnetic phase transition and it shows a sharp peak around $T_C$.[8] Our above results indicate that the high frequency *ac* magnetoresistance in our sample is sensitive to the intra-grain magnetization process rather than to the tunneling magnetoresistance. While the exact origin of the intrinsic colossal magnetoresistance is still elusive, the intrinsic *dc* magnetoresistance in single crystalline $La_{1-x}Sr_xMnO_3$ is shown to scale with the magnetization according to $\Delta\rho/\rho = C(M/M_s)^2$ where $M_s$ is saturation magnetization and $C$ measures the effective coupling between the $e_g$ conduction electron and $t_{2g}$ local spins.[8] Majumdar and Littlewood[18] proposed a phenomenological model which attributes the large magnetoresistance observed near $T_C$ in wide band width itinerant ferromagnets such as $La_{0.7}Sr_{0.3}MO_3$ (M = Co, Mn) to the suppression of magnetic fluctuations by an external magnetic field. In this model, $C \propto (1/2k_F\xi_0)^2$, where $k_F$ is the Fermi wave vector of an electron gas of density $n$, i.e., $k_F = (3\pi^2 n)^{1/3}$ and $\xi_o$ is the bare spin-correlation length.

As mentioned in the introduction, the flow of high frequency current in the sample is confined to a surface layer of thickness, $\delta = \sqrt{(2\rho)/(\omega\mu_0\mu_t)}$. From the classical theory of electromagnetism, electrical impedance of a current carrying slab with thickness $2a$ is $Z/R_{dc} = (jka)\coth(jka)$, where $k = (1-j)/\delta$ is the wave propagation constant and $R_{dc}$ is the *dc* resistance.[19] The skin depth is much larger than the thickness of the slab ($a/\delta << 1$) at low frequencies, and the above the above expression for the impedance is expanded to give $Z = R_{dc}[1+(4/45)(a/\delta)^4+2j(a/\delta)^2/3] \approx R_{dc}[1+j2(a/\delta)^2/3] = R_{dc}[1-j(a^2\omega\mu_0\mu_t)/(3\rho)]$, where we have only retained the term up to quadratic in $a/\delta$. If we consider the complex nature of the transverse permeability, $\mu_t = \mu_t'-j\mu_t''$, we get $R/R_{dc} = [1+ a^2\omega\mu_0\mu_t''/(3\rho)]$ and $X/R_{dc} = a^2\omega\mu_0\mu_t'/(3\rho)$ where $Z = R+jX$. So the *ac* resistance $R$ comprises of the magnetic dissipation characterized by $\mu_t''$ and $R_{dc}$. Generally, $\mu_t''$ is negligible at low frequencies, increases with increasing frequency and shows a peak at a characteristic frequency corresponding to the domain wall relaxation, whereas as $\mu_t'$ is higher at low frequencies and decreases rapidly as the domain wall relaxes.[5,20] At $f = 100$ kHz, the resistive part $R$ is dominated by the $R_{dc}$ since the contribution from the $\mu_t''$ is negligible and hence the $R$ decreases smoothly across the $T_C$. On the other hand, $X$ shows a rapid increase at $T_C$ because $X \propto \mu_t'$. As the frequency increases to $f = 500$ kHz, the magnetic loss exceeds the decrease in $R_{dc}$ due to the rapid increase in $\omega\mu_t''/\rho$ at $T_C$, which in turn leads to the observed anomaly. As the frequency increases further, $\delta$ becomes comparable or lower than the sample thickness ($2a = 3.6$ mm). For example, the non magnetic skin depth of the sample at 300 K is $\delta = 11.6$ mm at $f = 100$ kHz, but is 1.6 mm at $f = 5$ MHz. The skin depth will be further reduced in the ferromagnetic state because of the large initial transverse permeability. For example, the magnetic skin depth at 300 K is $\delta = 0.16$ mm at $f = 5$ MHz, if we assume $\mu_t = 100$. In this case of strong skin effect ($\delta \ll 2a$), we can write $Z = (1+j)\rho/\delta = \sqrt{(j\rho\omega\mu_0\mu_t)}$. In the paramagnetic state $\mu_t = 1$ and the behavior of the impedance is expected to follow the temperature dependence of the resistivity. However, the rapid increase of $\mu_t$ at $T_C$ leads to a minimum in the penetration depth and a rapid increase of $Z$ at $T_C$. Since $\mu_t(H_{dc}, f) = \mu_t'(H_{dc}, f) - j\mu_t''(H_{dc}, f)$ the impedance $Z \propto \left[\sqrt{\mu_R} + j\sqrt{\mu_L}\right]$, where $\mu_R = \sqrt{(\mu_t'^2 + \mu_t''^2)} + \mu_t''$ and $\mu_L = \sqrt{(\mu_t'^2 + \mu_t''^2)} - \mu_t''$, and the real part of the impedance represents the power absorption by the sample.[5] Hence, the extraordinary sensitivity of $R$ and $X$ at low fields can be understood as a result of the decrease of $\mu_R$ and $\mu_L$ by the $dc$ bias field. It is to be noted that while a $dc$ bias field of $\mu_0 H_{dc} = 1$ kOe completely suppressed the maximum in $R$ at $f = 500$ kHz, it did not eliminate the maximum at 1 and 2 MHz. Higher magnetic fields are needed to increase the skin depth further for $f \geq 1$ MHz.

The decrease in the amplitude and downward shift of $R$ and $X$ in temperature with increasing $dc$ bias field is reminiscent of the behavior of $ac$ susceptibility under increasing $dc$ bias magnetic field, as reported in a variety of systems including spin glass ($BiMn_{0.6}Sc_{0.4}O_3$)[21] and cluster glass[22] ($La_{0.67}Ca_{0.33}Mn_{1-x}Ta_xO_3$, ($x \geq 0.05$)) compounds as well as in a few long-range itinerant ferromagnets such as $ZrZn_2$,[23] $Pr_{0.5}Sr_{0.5}CoO_3$,[24] and $La_{0.73}Ba_{0.27}MnO_3$.[25] In a long range ferromagnet such as the tiled compound, the observed behavior is most likely connected with the dynamics of domain walls. Application of a *dc* bias field enlarges the domain size and aligns magnetization of a domain along its direction. Hence, the smaller *ac* magnetic field is unable to move the domain walls further. As a result, the amplitude of the *ac* susceptibility decreases with increasing bias field. Very close to $T_C$, the suppression of spin fluctuations further dramatically reduces the ac susceptibility which leads to an increase in the magnetic penetration depth, resulting in huge negative magnetoimpedance as observed. The position of the maximum in the temperature dependence of susceptibility is most likely determined by the competition between the thermal energy which enhances spin fluctuations and energies involved in pinning domain walls. Although the observed magnetoimpedance effect can be considered as an outcome of the classical skin effect, extending ac impedance to radio frequency range in phase separated manganites having both ferromagnetic metallic and paramagnetic or antiferromagnetic insulating phases will be worthy from the view point of understanding interaction between electrical and magnetic dipoles or magnetocapacitance effect.[26]

In summary, we have shown the occurrence of a low field *ac* magnetoresistance ($\Delta R/R$ = 45 % at $f$ = 2 MHz) and equally large magnetoreactance ($\Delta X/X$ = 40 % at $f$ = 0.1 MHz) at $\mu_0 H$ = 1 kOe around the Curie temperature in a polycrystalline $La_{0.67}Sr_{0.33}MnO_3$ through the *rf* impedance measurements. We have attributed our observations to the suppression of magnetic fluctuations near $T_C$ which causes increase in the magnetic penetration depth and decrease in the impedance. However, there are several questions still remain unanswered: What is the role of tunneling magnetoresistance at higher frequencies? What are the influences of microstructure on the magnetoimpedance? Can we observe the low-field *ac* magnetoresistance in single crystals or epitaxial thin films? Therefore, we believe that further investigations based on the present experimental results are essential to understand the fundamental physics of high frequency magnetotransport in oxides and exploit its technological applications as magnetic sensors.

R. M. acknowledges the National Research Foundation (Singapore) for supporting this work through the grant NRF-CRP-G-2007.

**Figure Captions:**

**Fig. 1** Temperature dependence of the *ac* resistance (*R*) and reactance (*X*) of $La_{0.7}Sr_{0.3}MnO_3$ for $f = 100$ kHz at $\mu_0H_{dc} = 0$ Oe. The downward pointed arrow indicates the onset of ferromagnetic transition determined from *ac* susceptibility.

**Fig. 2** Temperature dependence of the *ac* resistance (*R*) and reactance (*X*) of $La_{0.7}Sr_{0.3}MnO_3$ for $f = 100$ kHz and 500 kHz under different *dc* bias magnetic fields ($\mu_0H_{dc}$).

**Fig. 3** Temperature dependence of the *ac* resistance (*R*) and reactance (*X*) of $La_{0.7}Sr_{0.3}MnO_3$ for $f = 1$ MHz and 2 MHz under different *dc* bias magnetic fields ($\mu_0H_{dc}$).

**Fig. 4** Temperature dependence of the (a) *ac* magnetoresistance $\Delta R/R$ (%), (b) magnetoreactance $\Delta X/X$ (%) and (c) magnetoimpedance $\Delta Z/Z$ (%) under different bias magnetic fields ($\mu_0H_{dc}$) at $f = 2$ MHz. The insets show the frequency dependence of the peak value of the respective quantities at $\mu_0H_{dc} = 1$ kOe.

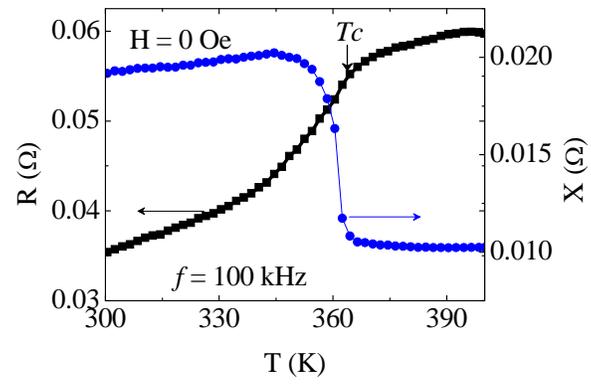

Fig. 1
Rebello *et al.*

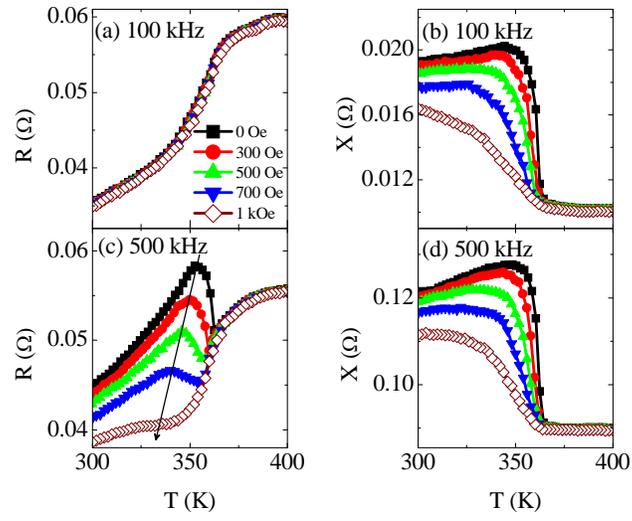

Fig. 2
Rebello *et al.*

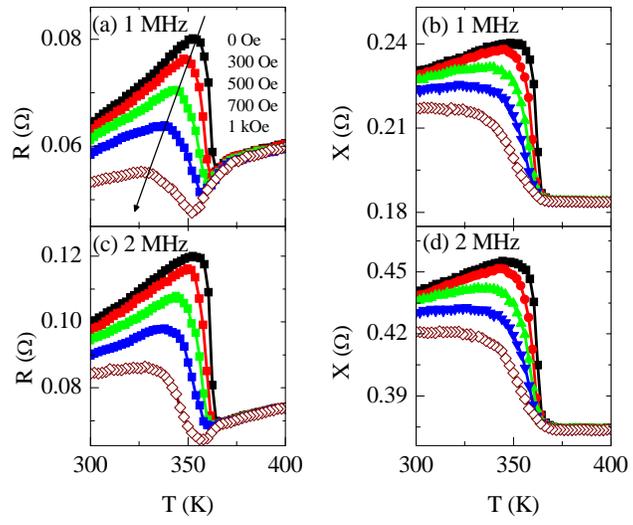

Fig. 3
Rebello *et al.*

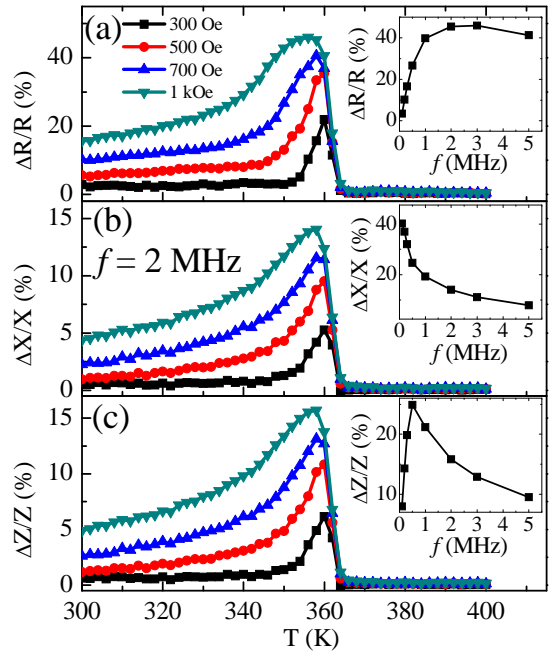

Fig. 4
Rebello *et al.*